# Effective chiral magnetic currents, topological magnetic charges, and microwave vortices in a cavity with an enclosed ferrite disk

Michael Sigalov, E.O. Kamenetskii, and Reuven Shavit

Department of Electrical and Computer Engineering,
Ben Gurion University of the Negev, Beer Sheva, 84105, Israel

**Abstract**

In microwaves, a TE-polarized rectangular-waveguide resonator with an inserted thin ferrite disk gives an example of a nonintegrable system. The interplay of reflection and transmission at the disk interfaces together with the material gyrotropy effect gives rise to whirlpool-like electromagnetic vortices in the proximity of the ferromagnetic resonance. Based on numerical simulation, we show that a character of microwave vortices in a cavity can be analyzed by means of consideration of equivalent magnetic currents. Maxwell equations allows introduction of a magnetic current as a source of the electromagnetic field. Specifically, we found that in such nonintegrable structures, magnetic gyrotropy and geometrical factors leads to the effect of symmetry breaking resulting in effective chiral magnetic currents and topological magnetic charges. As an intriguing fact, one can observe precessing behavior of the electric-dipole polarization inside a ferrite disk.

The concept of nonintegrable, i.e. path-dependent, phase factors is one of the fundamental concepts of electromagnetism. When there is no symmetry with rotational and/or translational invariance and so the wave equation cannot be separated in some coordinate system, one has an example of a nonintegrable system. Presently, nonintegrable systems (such, for example, as Sinai billiards) are the subject for intensive studies in microwave cavity experiments [1]. In view of the so-called quantum-classical correspondence, these experiments are useful in studying the quantum chaos phenomena. To get microwave billiards with broken time-reversal symmetry, ferrite samples were introduced into the resonators [2]. In our recent paper [3], we studied the microwave vortices in a three dimensional system of a TE-polarized rectangular-waveguide resonator with an inserted thin ferrite disk based on full Maxwell-equation numerical solutions of the problem. Because of inserting a piece of a magnetized ferrite into the resonator domain, a microwave resonator behaves under odd time-reversal symmetry (TRS) and a ferrite disk acts as a topological defect causing induced vortices. The microwave vortices are defined as lines to which the Poynting vector is tangential. The interplay of reflection and transmission at the disk interfaces together with material gyrotropy effect gives rise to a rich variety of wave phenomena. It was shown that the power-flow lines of the microwave-cavity field interacting with a ferrite disk, in the proximity of its ferromagnetic resonance, form the whirlpool-like electromagnetic vortices.

In different studies with TE polarized cavities [1 – 3], the vortex behavior in a vacuum region of a cavity can be easily understood from an analysis of the field structure. For TE polarized (with an electric field directed along *y* axis) electromagnetic waves in vacuum, the singular features of the complex electric field component $E_y(x,z)$ can be related to those that will subsequently appear in the associated two-dimensional time-averaged real-valued Poynting vector field $\vec{S}(x,z)$. For such electromagnetic fields, the Poynting vector is represented as $\langle\vec{S}\rangle = \frac{c^2}{8\pi\omega}\text{Im}(E_y^* \vec{\nabla}_\perp E_y)$, where $E_y$ is a complex vector of the *y*-component of the electric field: $E_y \equiv (E_c)_y e^{i\omega t}$. The fact that for electromagnetic fields invariant with respect to a certain coordinate, a time-average part of the Poynting vector can be approximated by a scalar wave function, allows analyzing the vortex phenomena. For a TE polarized field, we can write $E_y(x,z) \equiv \psi(x,z) = \rho(x,z) e^{i\chi(x,z)}$, where $\rho$ is an amplitude and $\chi$ is a phase of a scalar wave function $\psi$. This allows rewriting the Poynting vector expression as $\langle\vec{S}\rangle = \rho(x,z)^2 \nabla_\perp \chi(x,z)$. Such a representation of the Poynting vector in a quasi-two-dimensional system gives possibility to define a phase singularity as a point (*x*, *z*) where the amplitude $\rho$ is zero and hence the phase $\chi$ is undefined. These singular points of $E_y(x,z)$ correspond to vortices of the power flow $\langle\vec{S}\rangle$, around which the power flow circulates. A center is referred to as a (positive or negative) topological charge. Since such a center occur in free space without energy absorption, it is evident that $\nabla_\perp \cdot \langle\vec{S}\rangle = 0$. The singular points of $\langle\vec{S}\rangle$ (the vortex cores) can be directly related to the zero-electric-field topological features in a vacuum region of the cavity space, but not inside a ferrite region where one cannot express the Poynting vector only by the $\vec{E}$-field vector. Moreover, there is a special interest in an analysis of microwave vortices generated by a ferrite sample placed in a cavity region of a maximal field $E_y(x,z)$. As it was shown in [3], specifically the cases when a ferrite disk is placed in a maximum of the cavity electric field show the most pronounced and compact Poynting-vector vortices.

In the standard situation of microwave cavity experiments with inserted ferrite samples, the mechanism behind the TRS breaking effects is believed to be intimately connected with (and in fact generated by) the losses of energy and flux. Interplay between losses and quantum chaotic effects is rather interesting and non-trivial (see e.g. [4]). At the same time, it is known that the losses in the ferrite are always one of the main problems to study quantum manifestations of classical chaos in the regime of broken TRS. This fact makes a general comparison between the theory and experiment not an easy task. For this reason, any experimental technique able to generate TRS-breaking effects with introducing minimal losses (or introducing them in a controllable way) is of considerable interest.

In the ferromagnetic resonance, depending on a quantity of a bias magnetic field and (or) frequency, one has the regions with positive or negative permeability parameters [5]. For a ferrite disk placed in a maximum of the RF electric field in a rectangular-waveguide resonator, there are fundamentally different conditions for generation of microwave vortices in the positive- and negative-parameter regions. For negative permeability parameters, microwave vortices appear only when the material properties of a disk are characterized by big losses [6]. Similar situation takes place for a plasmon-resonance nanoparticle illuminated by the electromagnetic field, where the spiral energy flow line trajectories appear for a lossy sample with negative permittivity parameters [7]. Contrary, in a case of positive permeability parameters, the losses may play an indirect role in forming microwave vortices. This fact was illustrated in [3] (see Figs. 15 in [3]): in a case of positive permeability parameters one has very slight variation of the Poynting-vector vortex pictures in (and in close vicinity of) a ferrite disk for different losses parameters of a ferrite material.

The purpose of this letter is to show that for a ferrite disk with positive permeability parameters, creation of microwave vortices in a cavity is due to effective chiral magnetic currents rather than due to additional losses (which can be indeed minimal). As a very intriguing fact one can observe topological magnetic charges and precessing electric polarization in microwave ferrite samples, which appear because of the geometrical factor and the TRS breaking. Since the nonintegrable nature of the problem precludes exact analytical results for the eigenvalues and eigenfunctions, numerical approaches are required. We used the HFSS (the software based on FEM method produced by ANSOFT Company) CAD simulation programs for 3D numerical modeling of Maxwell equations [8]. In our numerical experiments, both modulus and phase of the fields are determined. We consider a situation when a normally magnetized ferrite disk (being oriented so that its axis is perpendicular to a wide wall of a waveguide) is placed in a region of a maximal cavity electric field in the middle of the waveguide height [3]. Fig.1 shows an example of such a configuration. The ferrite disk parameters are: diameter D=6 mm, thickness t=0.5 mm, saturation magnetization $4\pi M_o$=1880 Gauss, permittivity $\varepsilon_r$=15. A bias magnetic field of 5030 Oersted and the cavity resonance frequency of 8.7 GHz correspond to the region of positive permeability parameters. To stress the role of a combined effect of magnetic gyrotropy and geometrical factors we ignore any material losses in a ferrite disk (we assumed that $\Delta H$=0 and $\tan\delta$=0). The only losses taken into account in our numerical simulations are the losses in cavity walls: the cavity walls are made from copper.

A typical picture of the Poynting-vector microwave vortex [3] in air regions closely to a ferrite disk is shown in Fig. 2. In the center point, the Poynting vector is equal to zero. A ferrite disk is surrounded by an air cylinder and the Poyning vector is pictured on the upper and lower planes of this cylinder. Being aimed to uncover the physics of such a vortex structure, we should start with consideration of the electric and magnetic fields on the upper and lower planes of this external (air) cylinder. Fig. 3 shows the electric field vectors corresponding to a certain time phase, which we mark as phase $\omega t$=0º. At $\omega t$=180º one has an opposite direction of the electric field vectors. A thin ferrite disk slightly perturbs the cavity electric field. With variation of the time phase one has the 180°-switching of the electric field vector. At the same time, situation with the magnetic field distribution is completely different. Fig. 4 gives the top and side views of magnetic fields on the planes of the external (air) cylinder for different time phases. It becomes evident that for a certain point on the *xz* plane immediately above a ferrite disk, the magnetic field vector rotates in the clockwise direction. In Fig. 4, this can be viewed via rotation of the magnetic field vector in a given point A. The sizes of the in-plane magnetic field vectors reduce as one moves to a center.

The above results show that the main reason leading to creation of the Poynting vector vortices in the external (with respect to a ferrite) regions is strong spiral-like perturbation of the cavity magnetic fields caused by a ferrite disk. This perturbation can be described by effective magnetic currents. The notion of the effective magnetic currents is widely used in different excitation problems of waveguides and cavities [9]. There are equivalent surface magnetic currents which are formally introduced as a result of discontinuity of tangential components of an electric field: $\vec{i}^{\,m} \equiv \vec{n} \times (\vec{E}_{t_1} - \vec{E}_{t_2})$, where $\vec{E}_{t_1}$ and $\vec{E}_{t_2}$ are tangential fields near the interface and $\vec{n}$ is a normal to the interface. When we consider the fields inside a ferrite we find that on the upper and lower planes of a disk there are pronounced tangential components of the electric field vectors. In Fig. 5, the air cylinder has the same height as a ferrite disk and the picture shows the electric field distribution inside a ferrite disk and in the lateral-surface air vicinity of a ferrite disk. Comparing the pictures in Figs. 3 and 5 (above and below the surfaces of a ferrite disk), it becomes evident that transitions between the air-cylinder and ferrite-disk planes (the distances between these planes are much less than a waveguide height) can be formally considered as discontinuities of the tangential components of the electric field vectors. Fig. 6 shows top- and side-view time evolutions of the electric field in a ferrite disk. On the top-view pictures in Fig. 6, one sees the upper-plane vectors as intense arrows, while the lower-plane vectors are shown as pale arrows. The tangential electric field vectors on the lower plane are contrariwise to those on the upper plane. The in-plane electric field vector rotates in the counter clockwise direction. For the in-plane rotating clockwise magnetic field and counter clockwise electric field there is zero normal component of the Poynting vector in a ferrite disk.

The top-view pictures in Fig. 6 give clear perception of an effective surface magnetic current. In every given point of a ferrite-disk plane, an in-plane vector perpendicular to the tangential electric field is the vector of the equivalent magnetic current. Fig. 7 shows the surface magnetic current distributions on the upper plane of a ferrite disk. One can observe a spiral-like character of the surface magnetic current. Dynamics of the upper-plane-vector evolution shows that during a half of a time period a surface magnetic current flows away from a disk center and so one has a divergent spiral. During the next half of a period, a surface magnetic current flows towards to a disk center. There is a convergent spiral. Since the tangential-electric-field vectors on the upper and lower planes are in opposite directions, for the lower-plane-vector evolution (when one looks in the direction from the upper disk plane to the lower disk plane) there is a convergent spiral at the first half of the period and a divergent spiral at the second half of the period.

Dynamics of surface magnetic currents is correlated with the magnetic-field dynamics. The side-view pictures of the magnetic field distributions in Figs. 4 clearly show the presence of normal components of the magnetic field at certain time phases. More explicit pictures of the magnetic field distributions outside a ferrite disk (see Fig. 8) clearly demonstrate the convergent ($\omega t = 90°$) and divergent ($\omega t = 270°$) central locations of magnetic fields on a surface of a ferrite disk. A joint analysis of Figs. 7 and 8 gives an evidence for topological magnetic charges. The topological magnetic charges appear at the phases ($\omega t = 90°, 270°$) of extreme dynamical symmetry breaking when circulations of a surface magnetic current are maximal.

The main features of the microwave vortex creation appear from a special character of motion of the electric field vector inside a disk. It is worth noting that in a given point, the electric field inside a disk is characterized by a precessional behavior. An analysis shows that the electric-field polarization ellipse depends on the dielectric constant of the disk material. Fig. 9 (a) gives a picture of the in-plane electric-field polarization ellipse for a real ferrite material, having $\varepsilon_r$=15. In this case an axial ratio of the precession ellipse is 2.35. Fig. 9 (b) shows the in-plane polarization ellipse (in the same point of a disk) for a hypothetical ferrite material with $\varepsilon_r$=2. In this case an axial ratio of the precession ellipse is equal to 10.72. Because of a processional behavior of an internal electric field, electric-dipole polarization inside a ferrite disk should also have a precessional character of motion. This precessional motion of the electric-dipole polarization takes place against a background of the magnetization precession causing magnetic gyrotropy of the ferrite material. There is an evident coupling between two precessional motion processes, electric and magnetic.

In conclusion, we have to note that we found that in certain nonintegrable structures, magnetic gyrotropy and geometrical factors may lead to the effect of symmetry breaking resulting in effective chiral magnetic currents and topological magnetic charges. This is the main reason of creation of microwave vortices in a cavity with an enclosed lossless ferrite disk. We studied a normally magnetized thin-film ferrite sample. In this case of a quasi-2D structure, one has only in-plane components of magnetization, while the fields inside a ferrite are three dimensional. Following our previous results shown in paper [3], the vortex structure is strongly dependable on the disk position in a cavity. It is the subject of our future studies to analyze the effective magnetic currents and topological charges in different disk positions in a cavity.

One can expect opening a very exciting prospect in the electromagnetic-vortex applications. Recently, a new interesting phenomenon of relationships between the near-field phase singularities (vortices) in a slit-metal-plate structure and the features of the far-field radiation pattern has been observed [10]. The question of new types of microwave devices based on the considered above ferrite-disk vortex structures may appear as a subject of a great future interest. Some examples of proper applications of the vortex concept in design of the ferrite-based microwave devices can be found in [11]. It is shown, in particular, that for a microwave patch antenna with an enclosed ferrite disk the "vortex quality" is strongly correlated with the far-field antenna characteristics.

Figure captions
Fig. 1. (Color online) A rectangular-waveguide cavity with an enclosed ferrite disk
Fig. 2. (Color online) Top and side views of Poynting vector above and below a ferrite disk
Fig. 3. Electric field above and below a ferrite disk
Fig. 4. Top and side views of magnetic fields above and below a ferrite disk at different time phases
Fig. 5. (Color online) Electric field inside a disk and near a lateral surface of a disk
Fig. 6. (Color online) Top and side views of electric fields on the upper and lower planes inside a ferrite disk at different time phases
Fig. 7. Surface magnetic current distributions
Fig. 8. Magnetic field distributions outside a ferrite disk: Evidence for topological magnetic charges
Fig. 9. Tangential electric-field precession in a ferrite disk: (a) $\varepsilon_r = 15$, axial ratio of the precession ellipse is 2.35; (b) $\varepsilon_r = 2$, axial ratio of the precession ellipse is 10.72

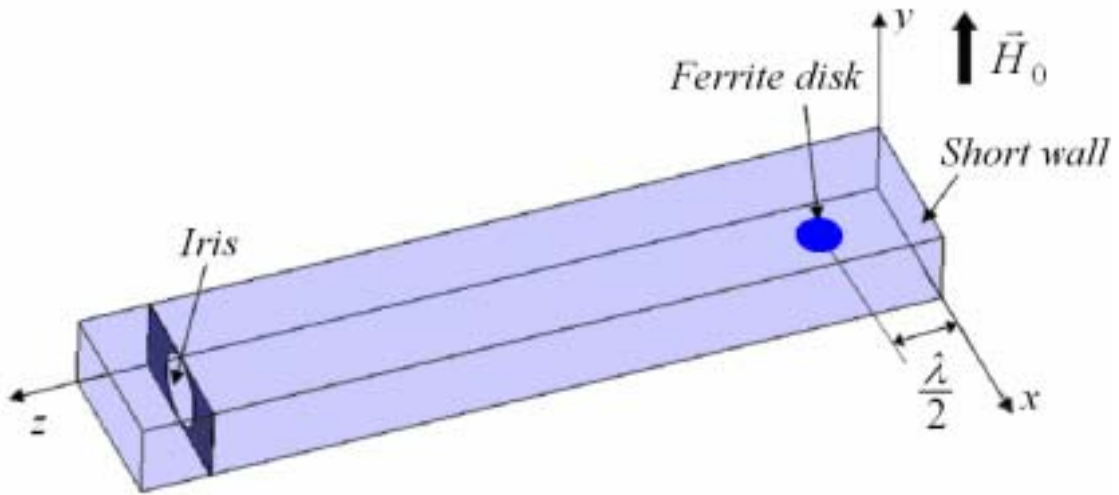

Fig. 1. (Color online) A rectangular-waveguide cavity with an enclosed ferrite disk

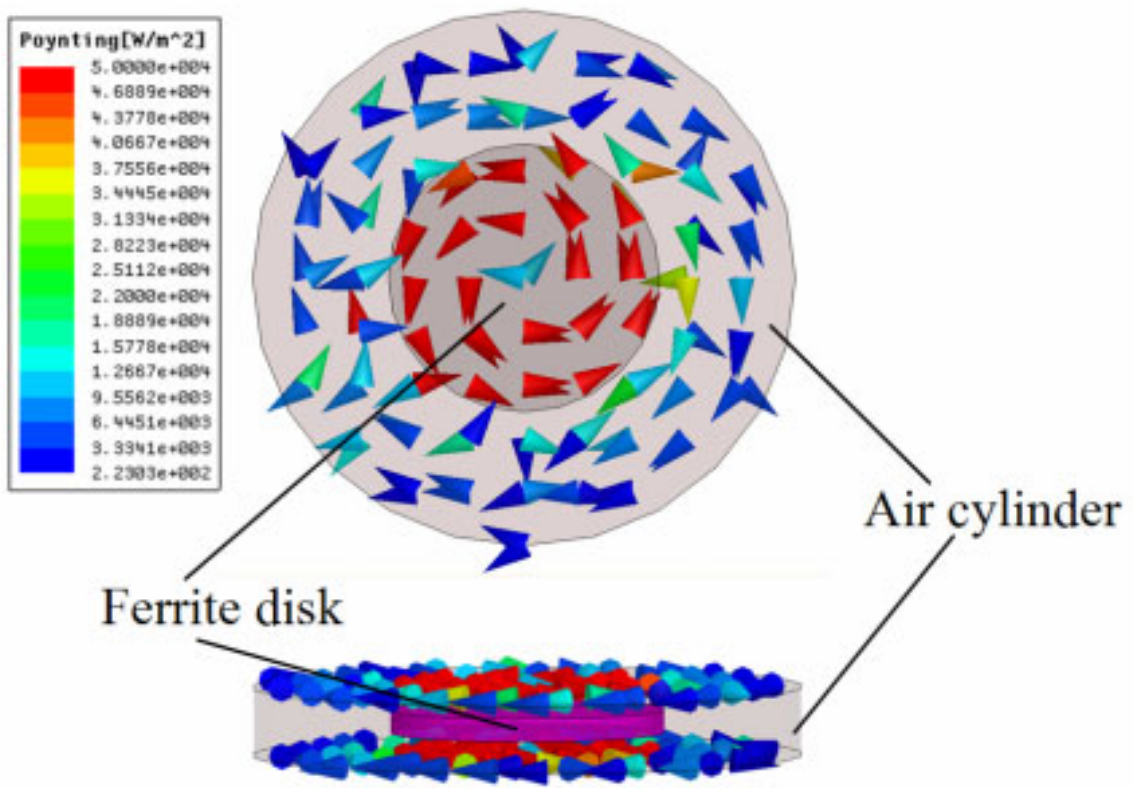

Fig. 2. (Color online) Top and side views of Poynting vector above and below a ferrite disk

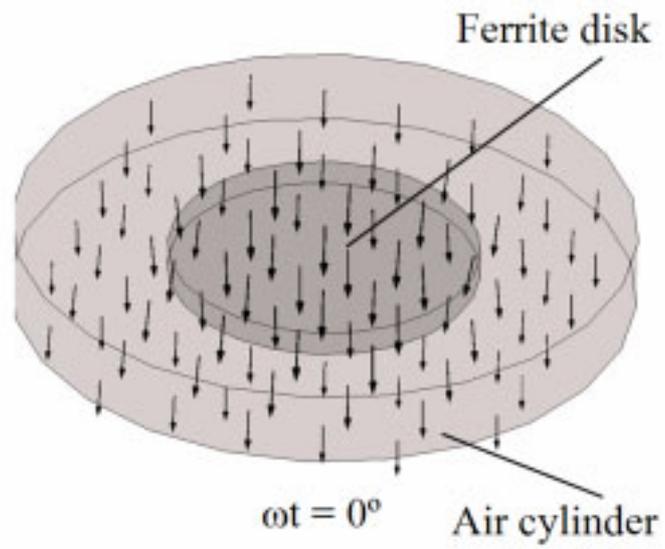

Fig. 3. Electric field above and below a ferrite disk at different time phases

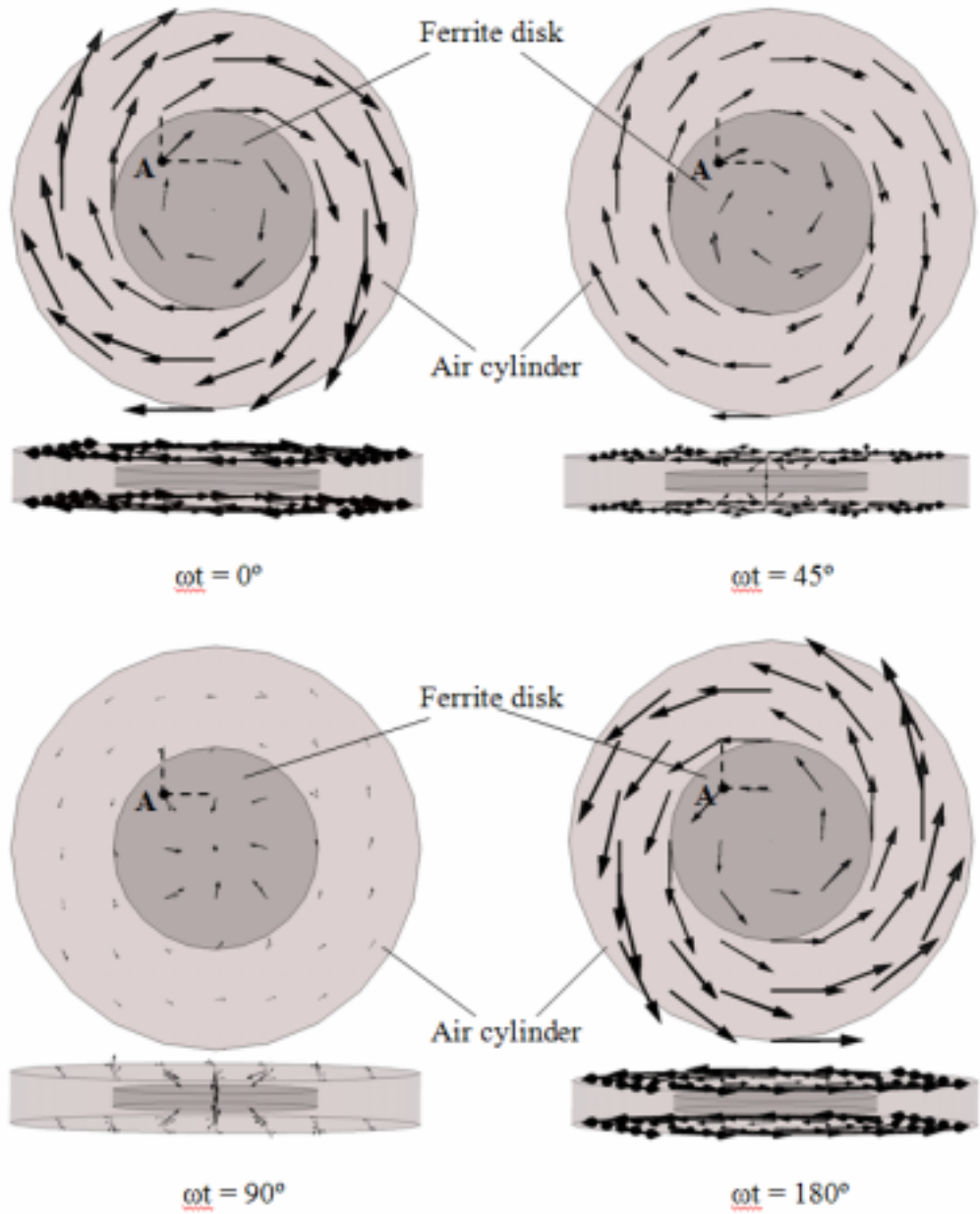

Fig. 4. Top and side views of magnetic fields above and below a ferrite disk at different time phases

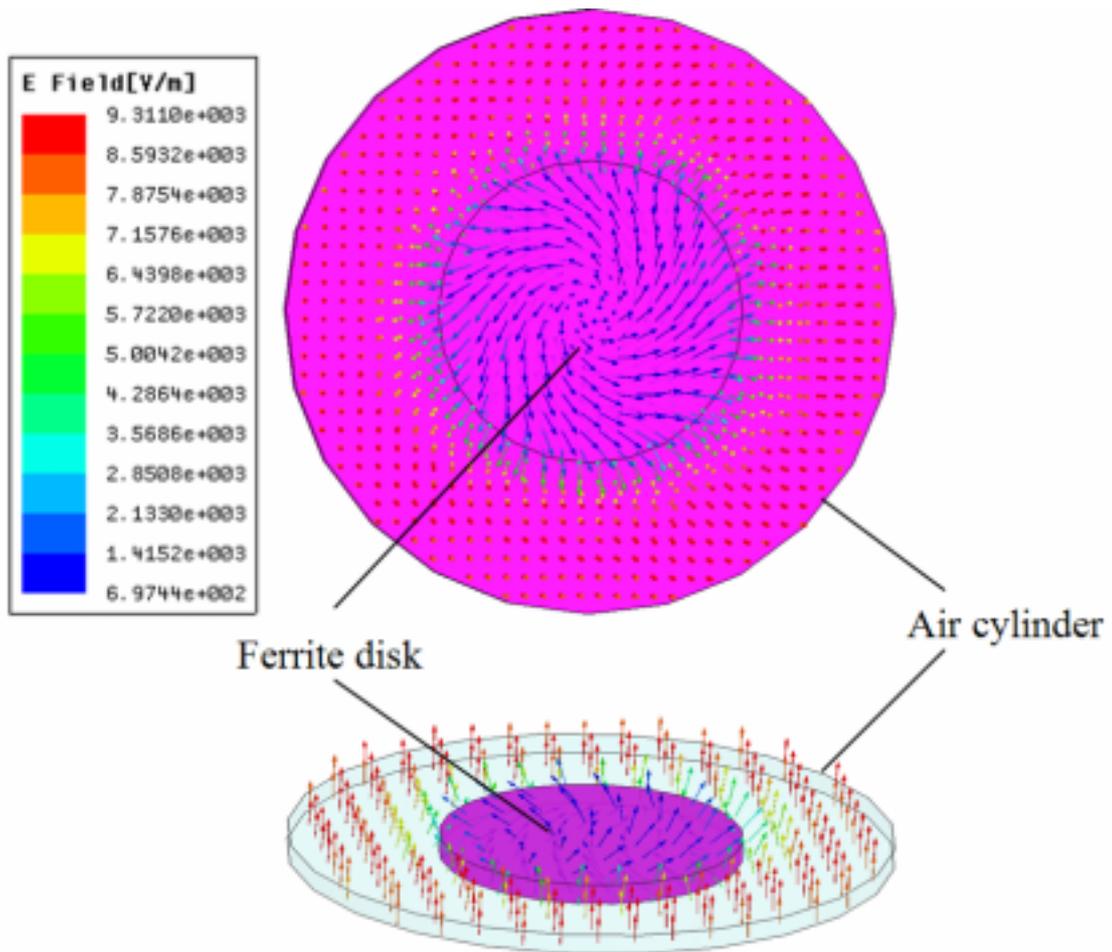

Fig. 5. (Color online) Electric field inside a disk and near a lateral surface of a disk

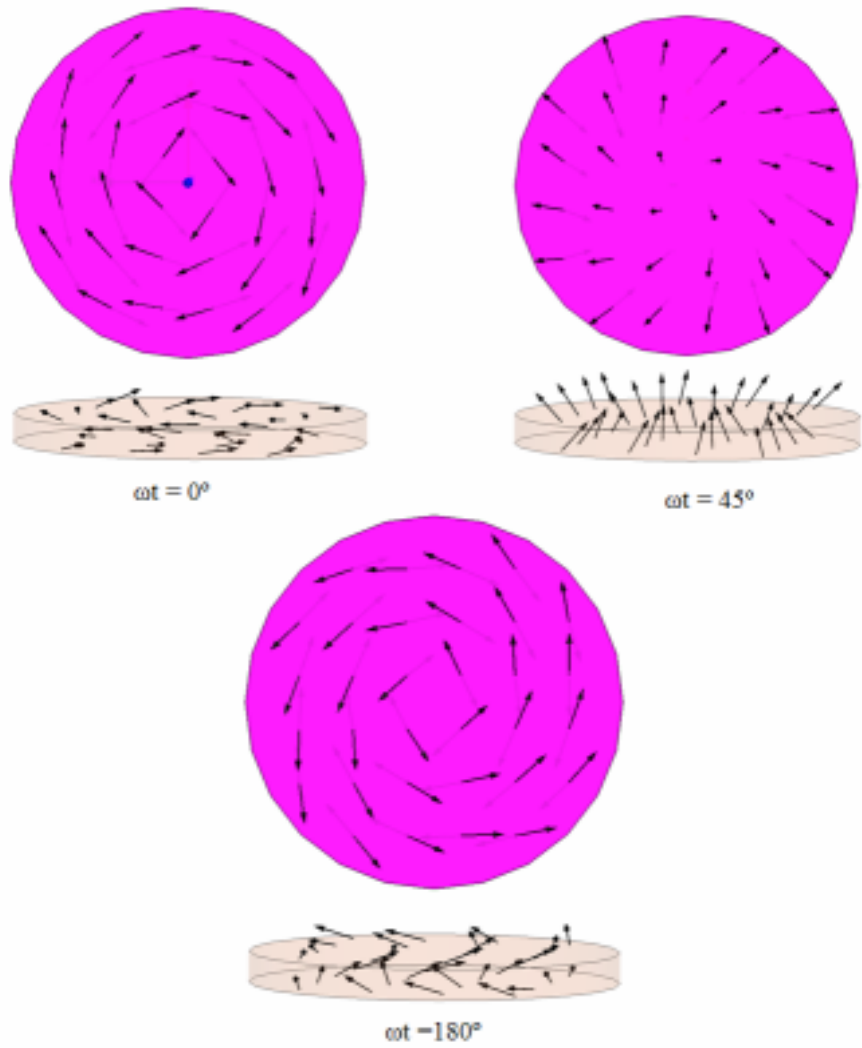

Fig. 6. (Color online) Top and side views of electric fields on the upper and lower planes inside a ferrite disk at different time phases

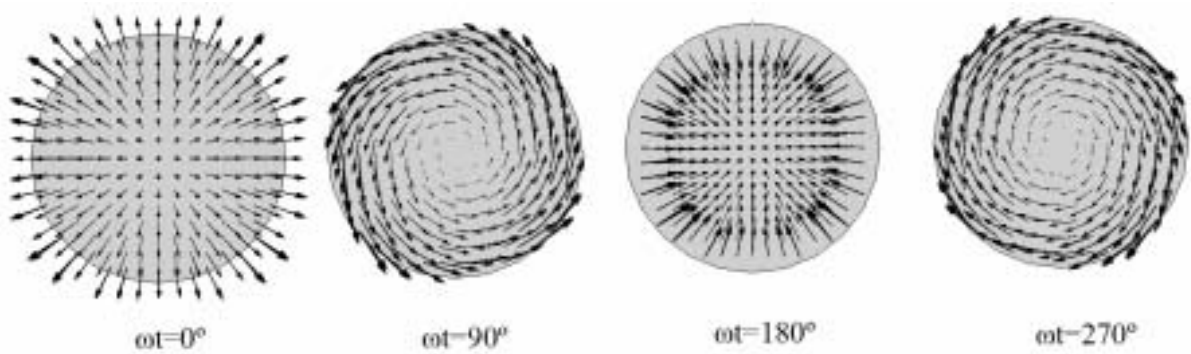

Fig. 7. Surface magnetic current distributions

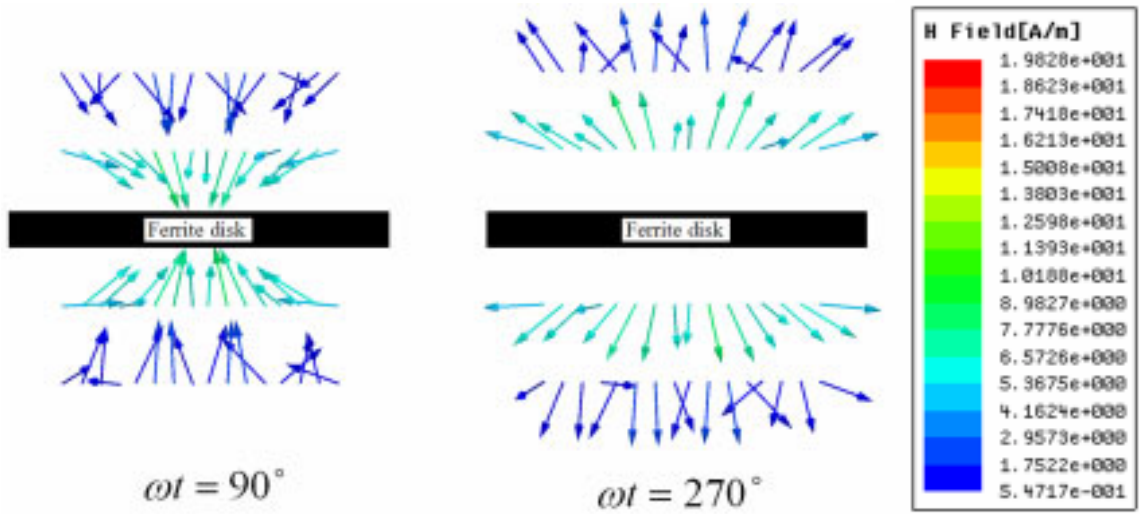

Fig. 8. Magnetic field distributions outside a ferrite disk: Evidence for topological magnetic charges

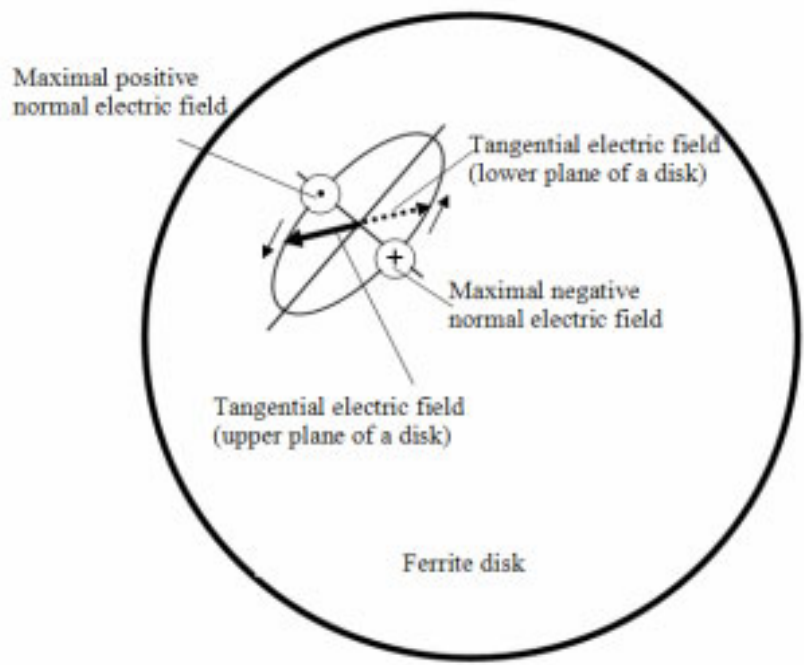

(a)

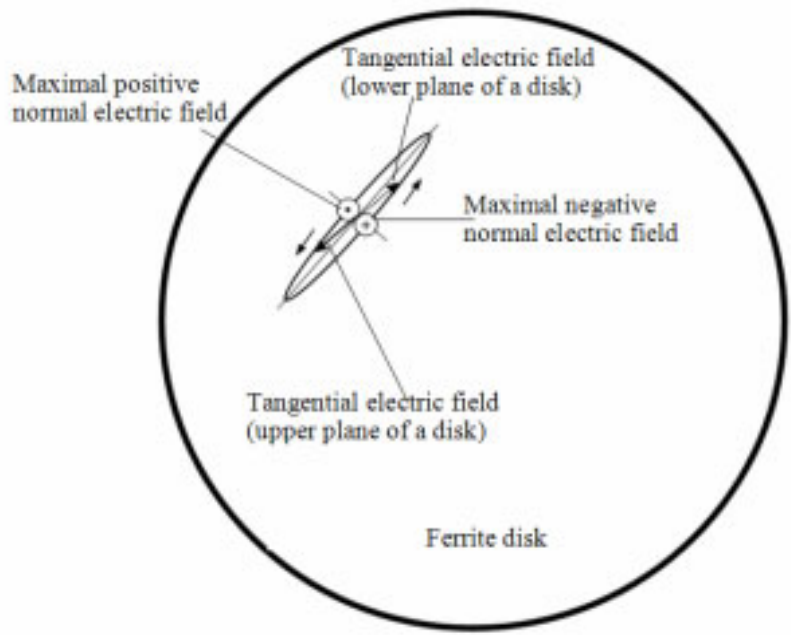

(b)

Fig. 9. Tangential electric-field precession in a ferrite disk: (a) $\varepsilon_r = 15$, axial ratio of the precession ellipse is 2.35; (b) $\varepsilon_r = 2$, axial ratio of the precession ellipse is 10.72